# Transport properties of dipole skyrmions in amorphous Fe/Gd multilayers

S. A. Montoya[1], M. V. Lubarda[2], V. Lomakin[1,3]

[1]*Center for Memory and Recording Research, University of California San Diego, La Jolla, CA 92093, USA*

[2]*Department of Mechanical and Aerospace Engineering, University of California San Diego, La Jolla, CA 92093, USA*

[3]*Department of Electrical and Computer Engineering, University of California, La Jolla, CA 92093, USA*

Dated: July 26, 2022

**Abstract**

The transport response of dipole skyrmions in amorphous centrosymmetric Fe/Gd multilayer is investigated by temperature and field-dependent resistivity measurements collected in three current and magnetic field configurations. It is shown that a dipole skyrmion lattice phase may form at certain temperatures leading to a unique signature of the polar longitudinal resistivity. This signature differs from the conventional field-varying parabolic response associated with stripe phases, which transition to a disordered skyrmion phase under applied fields. Transport measurements under different field history protocols reveal that the anomaly in the polar longitudinal resistivity appears under specific field history reversal processes. Our experimental results are reproduced using micromagnetic simulations that show the anomaly in the polar longitudinal resistivity is related to a domain wall reconfiguration that occurs as the domain morphology transitions from disordered stripes to a skyrmion lattice under applied perpendicular fields.

**Introduction**

Skyrmions, particle-like chiral magnetic spin textures, exhibit unique electric and magnetic properties given they carry quantized topological charge [1-2]. Numerous studies have shown that when conduction electrons interact with these non-trivial spin textures, the electrons couple to the magnetic texture of the skyrmion accumulating a Berry phase that gives rise to an additional Hall contribution, termed topological Hall effect [3-12]. Furthermore, magnetoresistance (MR) measurements reveal that a close-packed skyrmion lattice results in a unique signature in the MR-response [10-17]. In general, these electromagnetic properties have primarily been attributed to skyrmions which form in non-centrosymmetric magnets where the presence of a sizeable Dzyaloshinskii–Moriya interaction (DMI) favors a distinct canting between neighboring magnetic spins. Such canting ultimately embeds a single and global chirality among the formed magnetic spin textures. Since these novel electric and magnetic responses tend to result from electrons coupling with skyrmions, we seek of understand how the local and global chirality as well as their spatial arrangement contribute to the observed signatures. The single helicity of DMI skyrmions makes it difficult to discern chirality contributions. On the other hand, dipole stabilized skyrmions present a testbed to investigate these effects.

Here, we present a transport study of an amorphous Fe/Gd multilayer that exhibits magnetic phases ranging from stripe, coexisting stripe-skyrmion, close-packed skyrmion lattice, and disordered skyrmion [18-21]. These skyrmions possess a hybrid structure with a Bloch-line at the center of the film and opposite chirality Néel caps that extend from the center of the film toward the surface [18-21]. Given the lack of DMI, the skyrmion phase consists of an equal population of skyrmions with two opposite helicities: if the Bloch-line wraps continuously ($S = +1$) in a clockwise fashion then it has helicity $\gamma = -\pi/2$; similarly, a skyrmion with Bloch-line that wraps counter-clockwise is said to have helicity $= +\pi/2$ [18-21]. Previously, we demonstrated that these chiral spin textures are stabilized under a delicate balance of competing dipolar energy and domain wall energy [18-20], like achiral stripe and bubble domains ($S = 0$) [22-28], reason for which we term them dipole-stabilized skyrmions. Our experimental findings in centrosymmetric Fe/Gd multilayers show the magnetic phase transition from stripes to a close-packed skyrmion lattice results in a unique and field-dependent longitudinal resistivity anomaly, which differs from observations in non-centrosymmetric magnets where the helicity is preserved under applied magnetic fields [10-17]. By studying field history effects, we observe that the resistivity anomaly only emerges under specific field reversal conditions. Using numerical simulations, we validate our experimental transport observations and show that the longitudinal resistivity anomaly is correlated to a field-driven Bloch-line reconfiguration, which makes possible the formation of skyrmions under positive and negative magnetic fields in centrosymmetric Fe/Gd multilayers. Overall, our results demonstrate the mechanism by which dipole-stabilized skyrmions form under applied fields in centrosymmetric magnets, as well as show how dipole-stabilized magnetic stripes and skyrmions provide a platform to investigate the interplay between local and global chirality effects.

**Experimental Details**

The Fe/Gd films were prepared by DC magnetron sputtering by alternatively depositing Fe and Gd thin layers [18-21] at room temperature using an Argon process gas pressure of 3 mTorr in a chamber whose base pressure was less than 1 x $10^{-8}$ Torr. A seed and capping layer of 5 nm Ta is used to protect the film from corrosion. To study the transport properties of skyrmions in Fe/Gd multilayers, we chose an Fe and Gd layer composition that primarily favors a stripe phase that undergoes a transition to disordered skyrmion phase under applied perpendicular fields at/near room temperature, and a stripe phase that undergoes a field driven transition into a mixed stripe-skyrmion, close-packed skyrmion lattice, and disordered skyrmion phase at low temperatures [18-19]. The magnetic specimen exhibiting a close packed skyrmion lattice from 180K to 100K, of the form [Fe (0.36nm) /Gd (0.4nm)]x80, is deposited onto a photolithography defined 8-bar Hall cross transport device on Si substrate with a 300nm thermal oxide layer, and native-oxide Si substrate for magnetic characterization.

Transport measurements were performed using the standard 4-point probe AC (~10 Hz) bridge technique with an applied current of I = 1 µA. The patterned wires have a length of 3.5 mm and width of 10 µm. Field-dependent longitudinal resistivity ($\rho_{xx}$) and Hall resistivity ($\rho_{xy}$) measurements were collected as the magnetic field was swept from positive to negative saturation and conversely, at discrete temperatures from $T$ = 300 K to $T$ = 20 K in 20 K steps; meanwhile, temperature-dependent $\rho_{xx}$ measurements were collected at discrete temperatures from $T$ = 300 K to $T$ = 10 K in 2 K steps. We measure the $\rho_{xx}$ in three different magnetic field configurations: (i) in-plane and parallel to the length of the wire (parallel, $\rho_{xx}^{\parallel}$), (ii) in-plane and perpendicular to the length of the wire (transverse, $\rho_{xx}^{\perp}$), and (iii) perpendicular to the film (polar, $\rho_{xx}^{\odot}$). In our $\rho_{xx}$ and $\rho_{xy}$ measurements, the magnetic field is swept from $H$ = +10 kOe to $H$ = -10 kOe, and conversely.

**Results**

Figure 1(a) shows the field-dependent polar Hall resistivity ($\rho_{xy}^{\odot}$) obtained for the Fe/Gd patterned wire at $T$ = 300K. Inspecting the $\rho_{xy}^{\odot}$ loop, we can clearly observe signatures the Fe/Gd patterned wire exhibits a stripe phase at zero-field: net zero $\rho_{xy}^{\odot}$ is analogous to net zero magnetic moment which indicates an equal population of opposite perpendicular magnetic domains exists at remanence. The monotonic and nonlinear increase in $\rho_{xy}^{\odot}$ with applied perpendicular fields, from zero up to the saturated $\rho_{xy}^{\odot}$, exhibits a characteristic response commonly seen in magnetic materials hosting stripe phases (i.e., identifying the presence of skyrmion lattice or disordered skyrmion phases is not directly possible from field-dependent magnetometry nor $\rho_{xy}^{\odot}$ loops).

Figure 1(b) shows the temperature dependence of the polar longitudinal resistivity ($\rho_{xx}^{\odot}$) for the Fe/Gd patterned wire under a zero-field cooling protocol. The measurement shown in Fig. 1(b) were collected after applying a perpendicular field up to positive saturation and then reducing the field to zero-field. At $T = 300$ K, the average polar longitudinal resistivity is relatively high ($\rho_{xx}^{\odot} \sim 260\mu\Omega$-cm) in comparison to ferromagnetic alloys [29] which suggests the amorphous Fe and Gd layers are disordered and there is intermixing between the layers, as we have previously shown in Ref. [18]. As the temperature is reduced, from $T = 300$ K to $T = 10$ K, we observe $\rho_{xx}^{\odot}$ increases for the Fe/Gd patterned wire which suggests the specimen is semi-metallic; the latter observation is unexpected because most metals exhibit a drop in $\rho_{xx}^{\odot}$ as the temperature is reduced [29].

Figure 2 shows the field-dependent parallel ($\rho_{xx}^{\parallel}$), transverse ($\rho_{xx}^{\perp}$), and polar ($\rho_{xx}^{\odot}$) longitudinal resistivity at discrete temperatures, ranging from $T = 300$ K to $T = 80$ K in 20K steps, for the Fe/Gd patterned wire. In general, we know the longitudinal resistivity response depends on the scatter angle between the current and the magnetization. Lower scattering of electrons is expected from magnetic moments aligned perpendicular and orthogonal to the current flow, whereas moments parallel and anti-parallel to the current flow result in stronger scattering. Because the current is fixed along the length of the wire, different field orientations allow us to probe the behavior of different magnetic spin arrangements more favorably. Considering we know the Fe/Gd film possesses perpendicular magnetic spin textures with hybrid Néel-Bloch-Néel domain walls [18-19], we can deduce that: (i) $\rho_{xx}^{\odot}$ is more sensitive to scattering from magnetic spins in the Bloch-line and Néel caps, while (ii) $\rho_{xx}^{\parallel}$ and $\rho_{xx}^{\perp}$ are more sensitive to changes in the perpendicular component of magnetic spins.

By surveying the temperature dependence of the zero-field parallel ($\rho_{xx}^{\parallel}$), transverse ($\rho_{xx}^{\perp}$), and polar ($\rho_{xx}^{\odot}$) longitudinal resistivity for the Fe/Gd patterned wire, shown in Fig. 2, we can gain insights about the domain morphology and its 3-dimensional configuration. At room temperature ($T = 300$ K), both $\rho_{xx}^{\parallel}$ and $\rho_{xx}^{\perp}$ exhibit the same electron scattering from the ground state given $\rho_{xx}^{\parallel}(H = 0) = \rho_{xx}^{\perp}(H = 0)$, which is lower than the zero-field polar longitudinal resistivity, $\rho_{xx}^{\parallel}(H = 0) < \rho_{xx}^{\odot}(H = 0)$. Since we know the Fe/Gd multilayer exhibits perpendicular stripe domains with Néel caps and a Bloch-line [18-19], we can infer the stripe domains are disordered because the scattering at zero-field is the same along $\rho_{xx}^{\parallel}$ and $\rho_{xx}^{\perp}$ (i.e., aligned stripes would result in different electron scattering along parallel or transverse in-plane field geometries). Meanwhile, a higher zero-field $\rho_{xx}^{\odot}$ suggests the Néel caps occupy a sizable portion of the 3-dimensional magnetization volume fraction. As the temperature is reduced, the electron scattering from the ground state along $\rho_{xx}^{\parallel}$ and $\rho_{xx}^{\perp}$ both increase at different rates relative to each other, such that $\rho_{xx}^{\parallel}(H=0) > \rho_{xx}^{\perp}(H=0)$ below $T = 300$K, which indicates perpendicular stripes tend to order transverse the length of the Fe/Gd patterned wire. This rearrangement is possible when the magnetization of the Fe/Gd patterned wire is fully saturated and an applied in-plane field is reduced towards remanence [19]. Since $\rho_{xx}^{\parallel}(H=0) > \rho_{xx}^{\perp}(H=0)$, we can further infer the Néel caps occupy a larger volume fraction of the hybrid domain wall relative to the Bloch-line, and that Néel caps/Bloch line aligns parallel/transverse to the in-plane field [19]. Supplementary Section 1 details the anticipated longitudinal resistivity response and orientation of hybrid domain walls with different volume fraction [30].

Increasing the magnetic field, along either of the three transport configurations [Fig. 2(a), insert], results in longitudinal resistivity curves that are dependent on the domain morphology. We first focus on interpreting the temperature- and field-dependent $\rho_{xx}^{\parallel}$ and $\rho_{xx}^{\perp}$ response. Starting from zero-field, $\rho_{xx}^{\parallel}$ and $\rho_{xx}^{\perp}$ curves linearly change with increasing field. The observation of a symmetric $\rho_{xx}^{\parallel}$ and $\rho_{xx}^{\perp}$ responses, around zero-field, suggests that the field-dependent domain morphology evolves similarly along both positive and negative field directions. Moreover, the linear response indicates the domains smoothly and continuously tilt in the direction of the applied field. From $T = 300$ K to $T = 100$ K, $\rho_{xx}^{\parallel}$ and $\rho_{xx}^{\perp}$ exhibit a linear response that narrows around zero-field with decreasing temperature. Above the saturation field ($H_{sat}$), $\rho_{xx}^{\parallel}$ and $\rho_{xx}^{\perp}$ become constant because the magnetization is uniformly aligned in the direction of the applied field; hence, the scattering of conduction electrons does not change above magnetic saturation. The fact that we observe a reduction in the saturation field along $\rho_{xx}^{\parallel}$ and $\rho_{xx}^{\perp}$ implies tilting of the perpendicular magnetic spins in the direction of the applied in-plane field becomes easier as the temperature is lowered; this trend is possible because the Fe/Gd films exhibit a $Q$-factor, defined as the ratio between uniaxial anisotropy ($K_U$) and shape anisotropy ($2\pi M_s^2$), of less than 1 and an effective anisotropy ($K_{eff}$) that becomes more negative with decreasing temperature [18-19]. Below $T = 100$K, $\rho_{xx}^{\parallel}$ exhibits a very weak response between nonuniform and uniform (saturated) magnetic phases, while $\rho_{xx}^{\perp}$ swiftly changes under low in-plane fields.

Next, we focus on the evolution of the field-dependent polar longitudinal resistivity $(\rho_{xx}^{\odot})$ as a function of temperature, which shows the evidence that the domain morphology undergoes microstructure reconfiguration at/near temperatures at which a close-packed skyrmion lattice phase emerges [Fig. 2]. At room temperature ($T$ = 300K), $\rho_{xx}^{\odot}$ exhibits a parabolic field dependence where $\rho_{xx}^{\odot}$ smoothly decays around zero-field toward magnetic saturation [31-34]. The smooth decay of $\rho_{xx}^{\odot}$ indicates the stripe domains undergo a continuous evolution under applied magnetic fields, a common feature of this magnetic phase. The parabolic trend of $\rho_{xx}^{\odot}$ is mostly consistent from $T$ = 300 K to $T$ = 260 K, were changes in the $H_{sat}$ are mostly noted [Fig. 2(a-c)]. When the temperature is reduced, from $T$ = 240 K to $T$ = 220 K, $\rho_{xx}^{\odot}$ begins to exhibit a non-symmetric field reversal around zero-field [Fig. 2(d, e)]. We observe that when the field is swept from negative to positive magnetic saturation, $\rho_{xx}^{\odot}$ does not smoothly decay in a parabolic fashion, over a narrow field range, above zero-field [Fig. 2(e, d), insert]. Similar observations are also noted when the field is reversed from positive to negative magnetic saturation. By further reducing the temperature, from $T$ = 200 K to $T$ = 100 K, the prior kink in the field-dependent $\rho_{xx}^{\odot}$ becomes a well-defined anomaly [Fig. 2(f-k)], where $\rho_{xx}^{\odot}$ tends to locally increase over a narrow field region before it decays at higher magnetic fields in a parabolic trend. The local increase in $\rho_{xx}^{\odot}$ indicates higher electron scattering takes place, where one would typically expect reduced scattering, which could be explained by a local rearrangement of the 3-dimensional magnetic spin textures. Furthermore, we observe that the field range in which the $\rho_{xx}^{\odot}$ anomaly emerges tends to shift towards zero-field as the temperature is reduced. Below $T$ = 100 K, the amplitude of the $\rho_{xx}^{\odot}$ anomaly decays, which suggests that the energetics favoring the domain reconfigurations are no longer favorable [Fig. 2(l)].

Since the anomaly in the $\rho_{xx}^{\odot}$ is not symmetric with applied fields, we posit microstructural changes in the domain morphology are field history dependent. We performed $\rho_{xx}^{\odot}$ measurements on the Fe/Gd patterned wire under different field history protocols at two fixed temperatures, one which primarily favors stripe domains ($T$ = 300 K) and the other a close-packed skyrmion lattice ($T$ = 160 K) under applied fields, to gain insights on the origin of the $\rho_{xx}^{\odot}$ anomaly. The magnetic field history protocols are as follow: (i) First, the perpendicular field is set to positive magnetic saturation, and it is then reduced to negative magnetic saturation; afterwards, the field is swept from negative saturation to positive saturation. This is the field history which has been detailed for measurements shown in Figure 2. (ii) The second field history protocol proceeds (i) with the perpendicular field being reduced from positive saturation to zero-field, then increasing the field once again to positive saturation, and finally reducing it to zero-field. Figure 3(a, b) shows the field-dependent $\rho_{xx}^{\odot}$ at $T = 300$ K and $T = 160$ K, respectively, under both field history protocols. At room temperature, we observe that $\rho_{xx}^{\odot}$ exhibits a parabolic-like field response and there are no indications of any notable anomalies that are dependent on either field history protocol utilized [Fig. 3(a)]; meanwhile, at low temperatures, field-dependent effects in $\rho_{xx}^{\odot}$ can clearly be discerned. As before [Fig. 2(h)], the anomaly in $\rho_{xx}^{\odot}$ appears under field protocol (i) [Fig. 3(b)]. For field history (ii), when the field is swept from positive saturation to zero-field, the field-dependent $\rho_{xx}^{\odot}$ exhibits a parabolic-like trend that mimics the prior field response at high magnetic fields but differs at fields where the $\rho_{xx}^{\odot}$ anomaly was detected and also exhibits a lower $\rho_{xx}^{\odot}$ amplitude at magnetic fields preceding the anomaly. These differences suggest that the underlying magnetic spin textures and their ordering differ. Subsequentially increasing the field from zero-field to positive saturation, we observe $\rho_{xx}^{\odot}$ smoothly decays with a parabolic-trend and there is no emergence of the $\rho_{xx}^{\odot}$ anomaly. Lastly, sweeping the field from positive saturation to zero-field results in a field-dependent $\rho_{xx}^{\odot}$ response which mimics the prior sweep, thus the field-dependent evolution between magnetic phases is continuous. The transport responses under both field history protocols show that a discontinuous microstructure reconfiguration occurs when transitioning between magnetic close-packed skyrmion lattice phases is stabilized under opposite polarity fields.

To aid unraveling the source of the anomaly in the field-dependent $\rho_{xx}^{\odot}$, we performed micromagnetic simulations by solving the Landau-Lifshitz-Gilbert (LLG) equation using the FastMag micromagnetic simulator [35]. Using magnetic properties previously identified for these Fe/Gd films [18-19], we modeled the equilibrium states which form in a slab with volume of 10 μm x 10 μm x 80 nm which we discretize with 10 nm tetrahedra. To model a slab which supports a close-packed skyrmion lattice we used: $M_S = 400$ emu/cm$^3$, $K_U = 4 \times 10^5$ erg/cm$^3$ and $A = 5 \times 10^{-7}$ erg/cm. The equilibrium states on the slab were recorded as the magnetic field was swept from $H_z = -5000$ Oe to $H_z = +5000$ Oe in $H_z = 50$ Oe field steps with 10 ns rise and 20 ns relaxation time, in the high damping regime ($\alpha = 1$). Since the anisotropic magnetoresistance (AMR) ratio is relatively low for these films (~ 0.3%), we assume that scattering is uniform across the slab when computing the resistivity at each field step. The resistivity was calculated via:

$$\rho(\phi) = 1 + \Delta\rho \cos^2 \phi \text{ where } \Delta\rho = \frac{\rho_{||} - \rho_\perp}{\rho_\perp},$$

where $\phi$ is the angle between the magnetization and current direction. For our simulations, we set $\Delta\rho = 200\mu\Omega - cm$ based on results in Fig. 2 and assume that the current direction is along the two sides of the slab. The overall resistivity resulting from each equilibrium state is calculated as a network of resistors: elements lying in the direction of the current are added in series and elements transverse to the current direction are added in parallel. Figure 4(a) shows that the calculated field-dependent resistivity exhibits a parabolic-like dispersion with two major anomalies appearing when the field is swept from negative to positive saturation. We overlay the corresponding magnetic phases (disordered stripes [Fig. 4(b, c)], stripe-skyrmion transition [Fig. 4(d, e)], skyrmion lattice [Fig. 4(f, g)], disordered skyrmions [Fig. 4(h, i)], and uniform magnetization [Fig. 4(j, k)]) via color-shaded blocks on Fig. 4(a) to facilitate correlating signatures in the calculated resistivity to different magnetic phases. Supplementary Figure 2 shows the field-dependent evolution of the equilibrium states from negative to positive saturation [30]. The first anomaly is abrupt and occurs in a narrow negative field range when transitioning from disordered skyrmions into a close-packed skyrmion lattice. The second anomaly is broad and occurs over a large positive field range within the stripe-skyrmion transition phase. Inspecting the magnetic phases that form under opposite polarity magnetic fields [Figs. 4(b-k)], distinctions in size and shape of perpendicular domains and their domain walls can be noted (e.g., magnetic spin textures in the stripe, stripe-skyrmion, and skyrmion lattice phase are narrower under negative fields compared to those stabilized under the same positive fields), suggesting that the distribution of the in-plane versus perpendicular magnetic moments differs.

By examining the average magnetization ($<m_{x,y,z}>$) and average modulus magnetization ($<|m_{x,y,z}|>$), we can obtain insights how individual magnetization components and volume fraction profiles change with applied magnetic fields. Figures 5(a-c) show the field-dependent average modulus magnetization ($<|m_x|>, <|m_y|>, <|m_z|>$) obtained from processing field-dependent equilibrium states. As expected, the domain wall occupies the highest volume fraction at zero-field [Figs. 5(a, b)], while at positive/negative saturation all the magnetization is oriented along the negative/positive z-axis [Fig. 5(c)]. Sweeping the field from one polarity to another, each

$<|m_{x,y,z}|>$ profile exhibits a parabolic distribution which generally mimics the calculated field-dependent $\rho_{xx}^{\odot}$ [Fig. 4(a)]. Both $<|m_x|>$ and $<|m_y|>$ exhibit similar distribution with applied fields [Figs. 5(a, b)], which suggests the volume fraction of $m_x$ and $m_y$ tends to be uniform along the x- and y-axis throughout the domain morphology. In terms of the anomaly features noted in the calculated field-dependent resistivity [Fig. 4(a)]: the first anomaly, occurring under negative fields, exhibits a sharper/stronger response in $<|m_x|>$ and $<|m_y|>$ suggesting a more pronounced reconfiguration of the in-plane magnetization (i.e., domain wall) occurs compared to the perpendicular magnetization of the spin textures [Figs. 5(a-c)]. The second anomaly, occurring under positive fields, tends to show $<|m_{x,y,z}|>$ similarly changing when increasing the field from zero-field towards positive saturation. Figure 5(d-f) shows the field-dependent average magnetization ($<m_x>$, $<m_y>$, $<m_z>$) obtained from processing field-dependent equilibrium states. As the magnetic field is swept from negative to positive saturation, $<m_x>$ and $<m_y>$ profiles change polarity with increasing field several times [Fig. 5(d, e)] at/near field ranges where the anomalies appear [Figs. 4(a) and 5(a, b)]; while $<m_z>$ tends to monotonically increase with applied fields [Fig. 5(f)]. Changes in the magnetization polarity suggest the orientation and/or chirality of magnetic spin textures reconfigure with increasing fields.

To identify the microstructural changes in the domain morphology that cause $<m_x>$ and $<m_y>$ to change polarity, we surveyed the field-dependent evolution of the in-plane magnetization along the mid-height of the slab, where Bloch-lines are present. Figures 5(g-y) show field-dependent equilibrium states as the field is swept from negative to positive saturation. The images primarily depict the magnetization along the x-axis ($m_x$) at the mid-height of the slab ($z = 0$ nm), or the magnetization along the z-axis ($m_z$) at the top surface of the slab ($z = +40$ nm) for select fields. Near the negative critical field ($H_z = -2600$ Oe), a whole range of cylindrical-like spin textures with $S = -1, 0,$ and $+1$ chirality (i.e., antiskyrmions [36], bubbles, and skyrmions) are formed when the field is reduced below negative saturation [Fig. 5(g)]. Via field-driven elliptical instabilities [26], the cylindrical-like spin textures transform into dumbell-like spin textures retaining their original chirality [Figs. 5(h-i)]. Within/outside the field-of-view, spin textures with $S = 0$ and $S = -1$ tend to collapse or merge with neighboring magnetic spin textures, as the field is increased, and there is a predominance of $S = +1$ chiral spin textures with coexisting left and right helicities [Figs. 5(j, l)]. A chiral stripe is represfented by opposite helicity

(colored) Bloch-lines along the stripe domain wall (perimeter). As the field approaches zero-field, stripe domains primarily populate the domain morphology with some skyrmions still present [Figs. 5(m-o)]. Further increasing the field, a local reconfiguration and merger of chiral stripe and skyrmions occurs which results in a domain morphology primarily consisting of achiral stripe domains with many Bloch points [Fig. 5(o-q)]. Achiral stripe exhibit the same helicity (color) Bloch-lines on opposite stripe domain walls, while a Bloch point is represented by a white-colored feature on a domain wall where the helicity of the Bloch-line changes from left-to-right or vice versa along the length of the domain wall. The reconfiguration/merger of chiral-to-achiral spin textures results in a local change of chirality for the domain morphology, which takes places near/at fields at which the second anomaly is observed [Figs. 5(d, e)]. Also noted, under applied fields, is the local rerrangement of Bloch-lines within each stripe [Fig. 5(r-t)], which is possible due to the displacement of Bloch-points in the stripe domain wall. Increasing the field, achiral stripe domains pinch into cylindrical-like spin textures with $S = -1, 0,$ and $+1$ chirality, which corresopond to antiskyrmions [36], bubbles, and skyrmions, respectively [Figs. 5(u-v)]. Inspecting the stripe-to-skyrmion and skyrmions lattice phases, we find skyrmions predominantly occupy the domain morphology [Figs. 5(u-x)]. As the field approaches positive saturation, antiskyrmions [36] and bubbles tend to collapse first, while higher fields are required to collapse skyrmions [Fig. 5(y)]. Recently, we experimentally demonstrated the topological robustness of skyrmions versus bubbles in a similar Fe/Gd multilayer that can be tailored to favor either chiral or achiral phases under different field history protocols [37], which matches trends observed in our simulations. Overall, numerical simulations suggest that the field-dependent domain morphology undergoes a 3-dimensional local magnetization reconfiguration, which enables the formation of dipole skyrmions under opposite polarity applied fields.

**Discussion.**

To further elucidate competing local and global chirality effects observed in dipole-stabilized magnetic phases, we explore the anticipated perpendicular field transitions of global chiral and achiral magnetic phases. Figures 6(a-e) schematically details the Bloch-line evolution to support single-helicity skyrmions under opposite applied fields. If we assume the skyrmion lattice phase favors 100% right-handed ($R$) helicity skyrmions under negative fields [Fig. 6(a)], then reducing the field towards zero-field results in the elongation of skyrmions into stripe textures that possesses the same 100% right-handed ($R$) helicity [Fig. 6(b)]. Under negative fields, both positive and negative magnetization stripe textures possess 100% right-handed (*R*) Bloch helicity [Fig. 6(a, b)]. As the field approaches zero, we can expect that the widths of the opposite polarity stripe textures become equivalent [Fig. 6(c)]. Then, under positive applied fields, the positive/negative polarity perpendicular stripe textures grow/shrink with increasing fields [Fig. 6(d)] until the shrinking stripes pinch into a collection of 100% right-handed (*R*) helicity skyrmions [Fig. 6(e)] (i.e., a perpendicular stripe domain can also coalesce into a skyrmion). The detailed field evolution of 100% right-handed (*R*) Bloch helicity skyrmion lattice phase, from negative to positive applied fields, demonstrates the Bloch-line helicity is globally and locally conserved. Shifting our focus to achiral global phases, where we anticipate local chirality under opposite applied fields, we can similarly explore the field evolution of chiral spin textures with 50% left-handed (*L*) and 50% right-handed (*R*) Bloch helicity. Figures 6(f-j) illustrate a potential global achiral ground state and an anticipated Bloch-line evolution under applied fields. As before, similar helicity skyrmions elongate into stripe textures that preserve the helicity distribution [Figs. 6(f, g)]. Under negative and zero fields, the positive magnetization stripes possess 50% left-handed (*L*) and 50% right-handed (*R*) Bloch helicity, while negative magnetization stripes exhbit 50% left-handed (*L*) and 50% achiral (*A*) Bloch helicity [Fig. 6(g, h)]. To conserve the local chirality under high positive fields, it is clear that at least one Bloch-line rotation [Fig. 6(i)] is required to achieve a skyrmion lattice phase consisting of 50% left-handed (*L*) and 50% right-handed (*R*) Bloch helicity [Fig. 6(j)]; alternatively, four Bloch line rotations would be required to conserve the local chirality at positive fields [Supp. Fig. 3, Ref. 30]. Given the nature of dipole-stabilized magnetic spin textures with local chirality, there are other potential global achiral ground states, which can also exhibit 50% left-handed (*L*) and 50% right-handed (*R*) Bloch helicity, each possessing distinct Bloch-line rotations to conserve local chirality under applied fields. Moreover, multiple global achiral ground states can coexist and locally neighbor one another, which increases the complexity of magnetization dynamics when magnetic phases undergo field-induced morphological changes. Supplementary Figure 4 details other potential global achiral ground states, which possess distinct Bloch-line distribution and their anticipated Bloch-line reorientation to preserve their local chirality [30]. Altogether, this empirical analysis demonstrates that in order to properly understand and describe local and global chirality effects in centrosymmetric magnets there is a need to consider the helicity of all magnetic spin textures stabilized at any field including their spatial arrangment to understand their local and global electrical and magnetic responses.

**Conclusions.**

We have measured the transport properties of dipole skyrmions that form in Fe/Gd multilayer patterned wires through temperature- and field-dependent resistivity studies. By experimentally studying an Fe/Gd pattened wire with composition that primarily favors stripes at room temperature and a close-packed skyrmion lattice at low temperatures, we profiled the evolution of the transport signatures as a function of temperature and applied fields for three current-field geometries (parallel, transverse, and polar), which provided valuable insights about the dipole-stabilized magnetic spin textures and their 3-dimensional arrangement. First, we confirmed stripe phases undergo continuos morphological transformations under applied fields, which are characterized by a representative parabolic-like resistivity response under perpendicular fields, while skyrmion lattice exhibit a parabolic-like resistivity response with a nontrivial anomaly. Field history studies revealed that the presence of a close-packed skyrmion lattice results in discontinuos microstructural changes in the domain morphology when transitioning between magnetic phases stabilized under opposite magnetic fields. Through micromagnetic simulations, we reproduced our experimental observations of an anomaly in the field-dependent polar longitudinal resistivity, and obtained detailed insights on the field-dependent evolution of complex 3-dimensional magnetic spin textures. Micromagnetic simulations showed that the stripe phase undergoes a Bloch-line reorientation as the magnetic field is swept from one polarity to another, which enables the formation of an equal population of two helicity dipole skyrmions under negative and positive magnetic fields. Our results provide a strategy to aid the identification of materials capable of hosting dipole-stabilized skyrmion lattice phases, elucidate the mechanism by which dipole-stabilized skyrmions form in centrosymmetric magnets under applied perpendicular fields, and demonstrate how dipole-stabilized magnetic phases provide a distinctive testbed for studying interactions between local and global chirality effects.

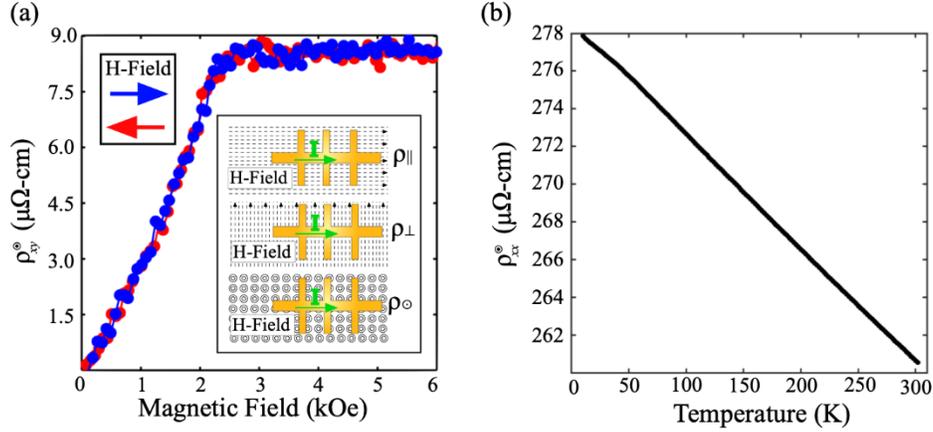

Figure 1. (a) Field-dependent polar Hall resistivity ($\rho_{xy}^{\odot}$) obtained at room temperature for the 10μm-wide Fe/Gd patterned wire. The $\rho_{xy}^{\odot}$ is color-coded to show field-dependent responses: (red-colored) positive-to-negative saturation, while (blue-colored) negative-to-positive saturation. The magnetic field was swept from $H_z = \pm 10$ kOe, but for ease of viewing a smaller field range is shown. The inset in (a) shows the magnetic field and current configurations used to measure the response of the Fe/Gd patterned wire. (b) The temperature-dependent polar longitudinal resistivity ($\rho_{xx}^{\odot}$) of the Fe/Gd patterned wire under a zero-field-cooling (ZFC) protocol; prior to field-cooling, the Fe/Gd patterned wire was saturated with a perpendicular field and then the field was reduced to $H_z = 0$.

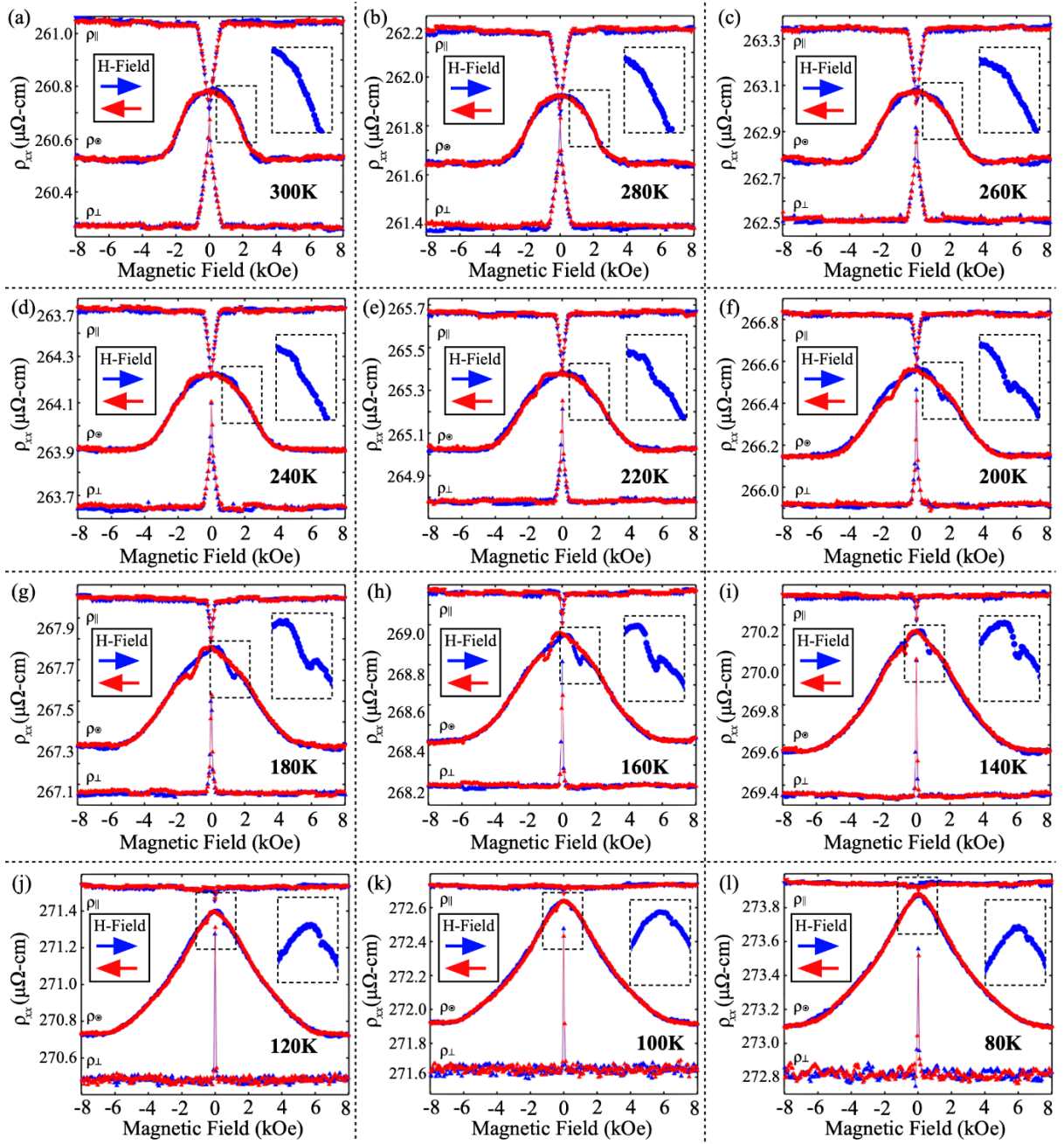

Figure 2. Field-dependent parallel ($\rho_{xx}^{\parallel}$), transverse ($\rho_{xx}^{\perp}$), and polar ($\rho_{xx}^{\odot}$) resistivity of the 10μm-wide Fe/Gd patterned wire from T = 300 K to T = 80 K in 20 K intervals. At each temperature, the magnetic field is swept from $H = \pm 10$ kOe, but for ease of viewing a smaller field range is shown. The $\rho_{xx}$ is color-coded to show field-dependent responses: (red-colored) positive-to-negative saturation, while (blue-colored) negative-to-positive saturation. The insets, at each temperature, show an enlarged view of features observed in $\rho_{xx}^{\odot}$ as fields are swept from negative to positive saturation.

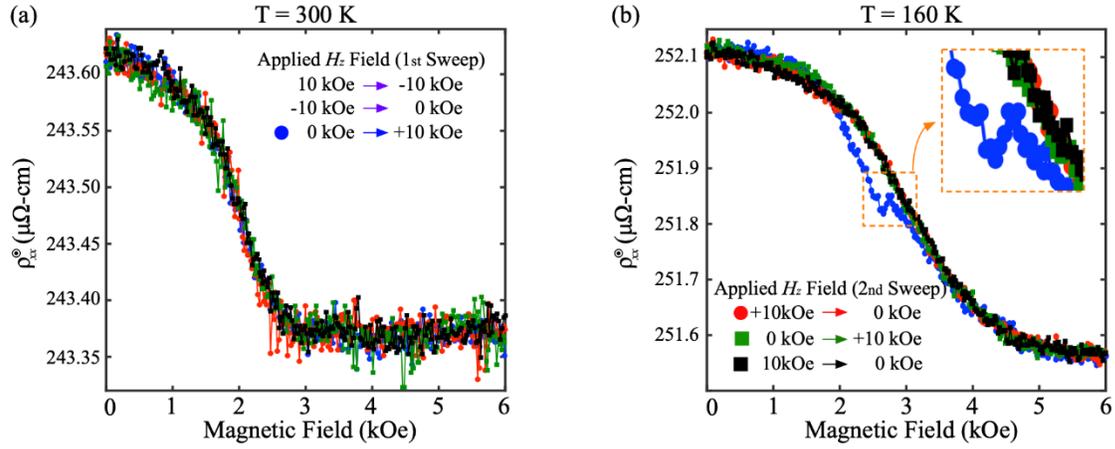

Figure 3. Field-dependent polar longitudinal resistivity ($\rho_{xx}^{\odot}$) of the 10μm-wide Fe/Gd patterned wire at (a) T = 300 K and (b) T = 160 K under two consecutive field history protocols. The insert in (a, b) show the $H_z$ field sweep protocols for the four field-dependent $\rho_{xx}^{\odot}$ responses: the first sweep is detailed in (a), while the subsequent field sweep is shown in (b).

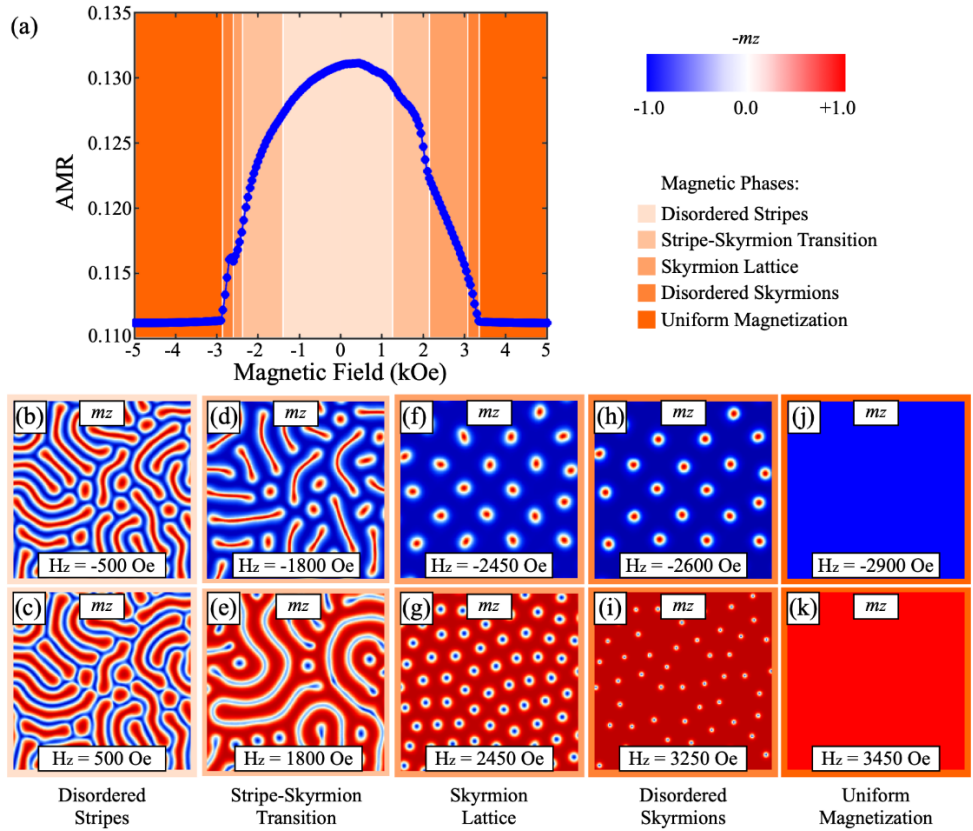

**Figure 4.** (a) Field-dependent calculated resistivity obtained from processing equilibrium states stabilized on a 10 $\mu m$ x 10 $\mu m$ x 80 $nm$ slab with 10 $nm$ tetrahedra, and magnetic properties: $M_S$ = 400 emu/cm³, $K_U$ = 4x10⁵ erg/cm³, and $A_{ex}$ = 5x10⁻⁷ erg/cm. Overlaid on (a) are the respective field-dependent magnetic phases (disordered stripes, stripe-skyrmion transition, skyrmion lattice, disordered skyrmion, uniform magnetization). (b-k) Equilibrium states, at different perpendicular fields, illustrate the top-side view of the magnetization along the z-axis ($m_z$) at the top surface of the slab ($z = 40\ nm$). The perpendicular magnetization ($m_z$) is represented by regions in red ($+m_z$) and blue ($-m_z$), while the in-plane magnetization ($m_x, m_z$) is represented by white regions. Each image depicts a 2 $\mu m$ x 2 $\mu m$ field of view near the center of the 10 $\mu m$ x 10 $\mu m$ total area. The colorbar details the ($m_z$) magnetization polarity detailed in the equilibrium states.

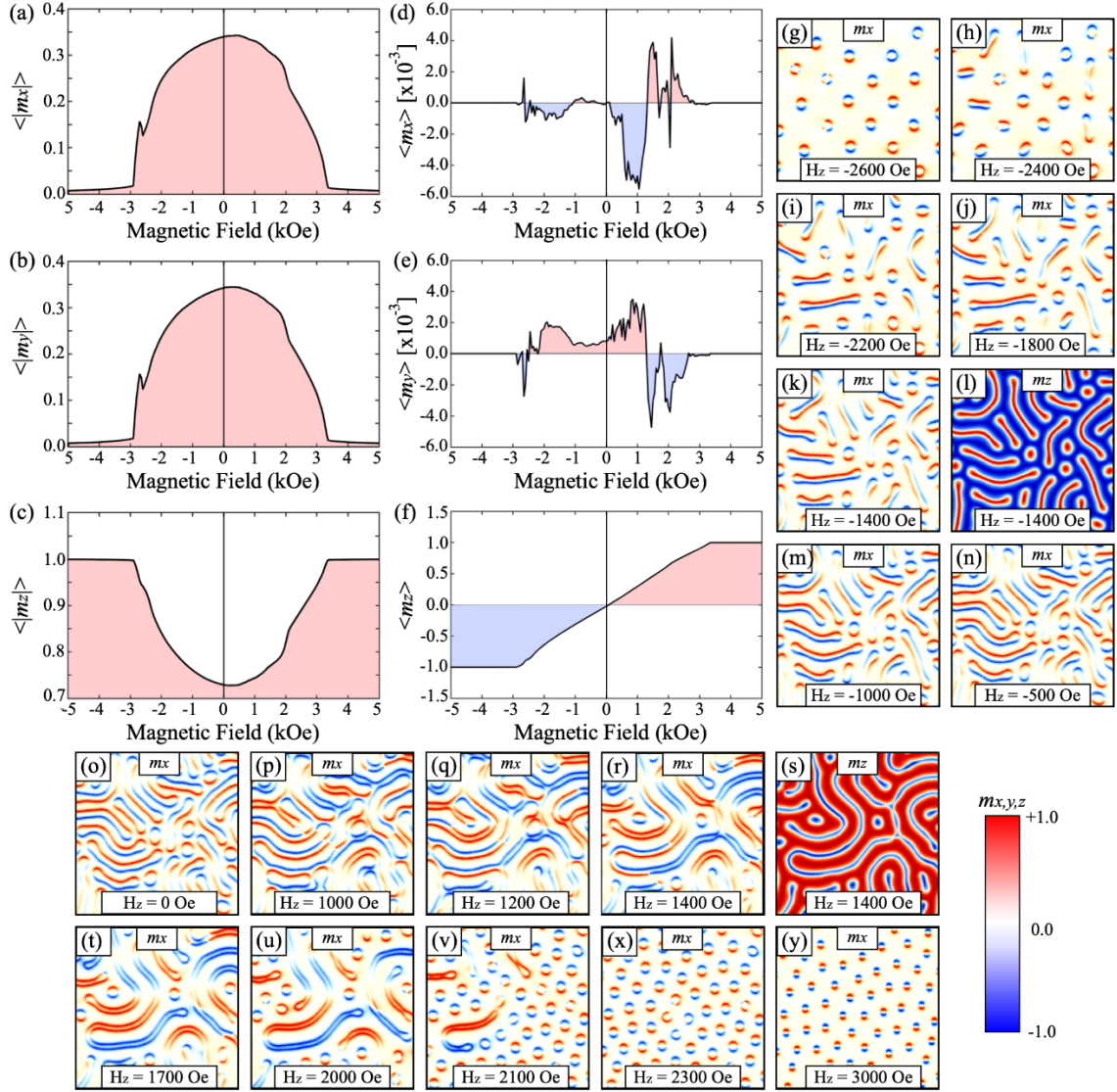

**Figure 5.** (a-c) The field-dependent average modulus of ($m_x$, $m_y$, $m_z$) magnetization. (d-f) The field-dependent average of ($m_x$, $m_y$, $m_z$) magnetization. Overlaid on (a-f) are reference-colors detailing red (positive) and blue (negative) magnetization. (g-y) Field-dependent evolution of equilibrium states as perpendicular field is swept from negative to positive saturation. Each image shows the top-side view of the magnetization along the x-axis ($m_x$) at the center of the slab ($z = 0\ nm$), or z-axis ($m_z$) at the top surface of the slab ($z = 40\ nm$). All the images capture the evolution of domain states in the same 2 $\mu m$ x 2 $\mu m$ region near the center of the 10 $\mu m$ x 10 $\mu m$ total area. The colorbar details for respective polarity of the ($m_x, m_z$) magnetization detailed in the equilibrium states. A reference for the eye is placed in Figs. 5(a-c) at $H_z = 0$ Oe to aid distinguishing $<|m_{x,y,z}|>$ features between negative and positive fields.

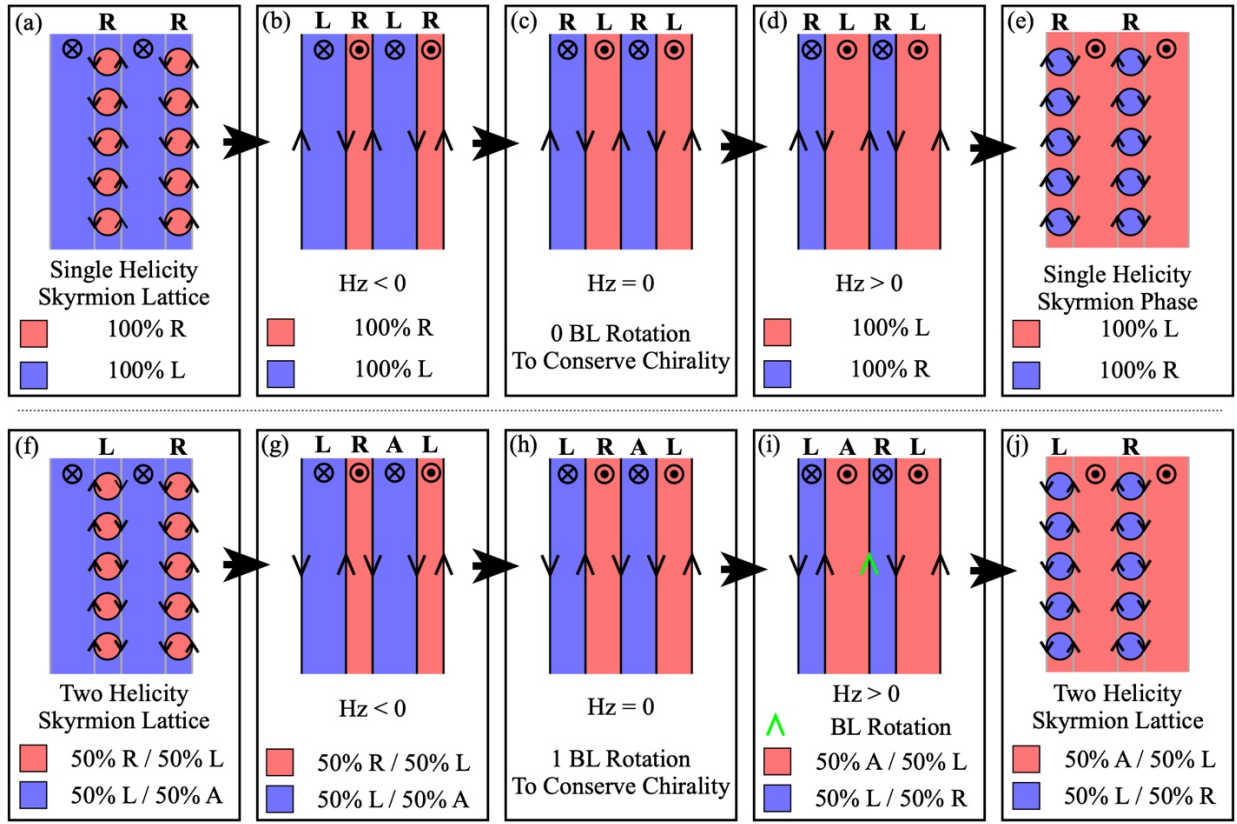

Figure 6. Schematic illustration of the Block-line evolution for global (a-e) chiral and (f-j) achiral skyrmion lattice phases as perpendicular magnetic field is applied from negative to positive (left to right images) saturation. The right-handed (R) and left-handed (L) Bloch-line helicity is depicted with a black-hue arrow-heads along the perimeter of red/blue-colored features which represent positive/negative perpendicular magnetic domains. A green-hue arrow head depicts a Bloch-line whose helicity changed compared to prior applied fields. For each ground state, the local and global Bloch-line helicity distribution is depicted for both perpendicular magnetization textures. Local/global Bloch-line helicity is shown above/below the spin texture arrangement.

# Supplementary Information

## Transport properties of dipole skyrmions in amorphous Fe/Gd multilayers

S. A. Montoya[1], M. V. Lubarda[2], V. Lomakin[1,3]

[1]Center for Memory and Recording Research, University of California San Diego, La Jolla, CA 92093, USA

[2]Department of Mechanical and Aerospace Engineering, University of California San Diego, La Jolla, CA 92093, USA

[3]Department of Electrical and Computer Engineering, University of California, La Jolla, CA 92093, USA

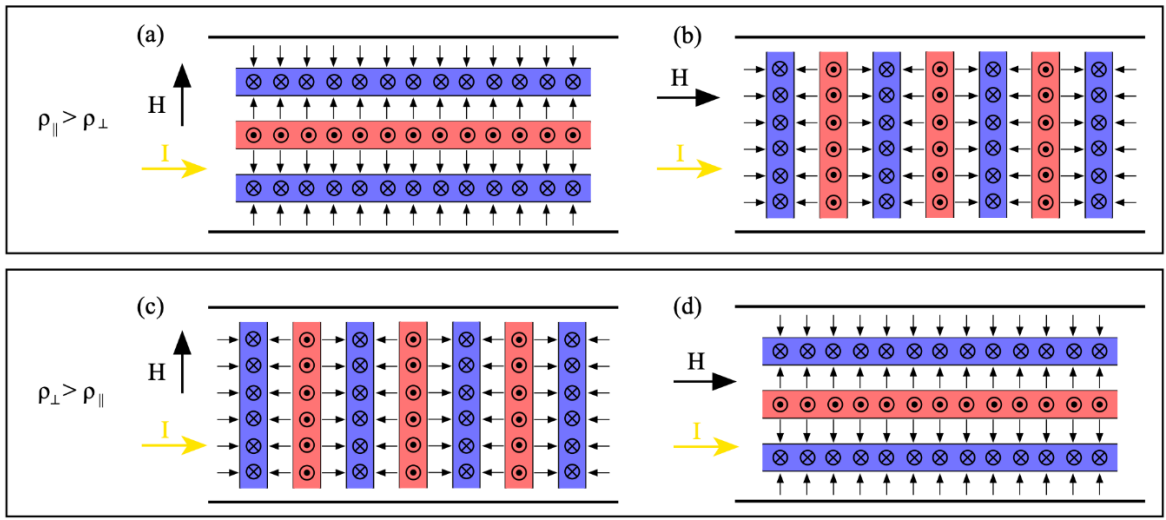

Supplementary Figure 1. Arrangement of perpendicular magnetic spin textures with hybrid domain walls under in-plane applied magnetic fields for a thick-film centrosymmetric magnet with material (a, b) $Q$-factor = 0.4 and (c, d) $Q$-factor = 0.6. Schematics detail the top-down view, along the top surface of a magnetic patterned wire, where red/blue-colored features represent positive/negative perpendicular magnetic domains, and the white-colored features detail Néel caps with orientation indicated by the overlaid, black-colored arrows. The horizontal black lines bounding each of the four schematics indicate the long axis direction of the patterned wire. Shown are the domain morphology arrangements under applied in-plane fields that are (a, c) transverse or (b, d) parallel to the long axis of the magnetic patterned wire.

1. *Arrangement of hybrid domain walls under in-plane magnetic fields in thick-film centrosymmetric magnet with material Q-factor <1.*

Supplementary Figure 1 shows the anticipated local/global arrangement of perpendicular magnetic spin textures with a hybrid domain wall under applied in-plane magnetic fields. The top-down view schematic representations primarily show the $+m_z$ magnetization (red-hue-colored features) and $-m_z$ magnetization (blue-hue-colored features) stripes that are separated by Néel caps (white-colored features) that would be observed at the top-surface of a magnetic slab. Previously [1], we demonstrated Bloch-line and Néel caps of hybrid domain walls that form in a thick-film centrosymmetric magnet with material $Q$-factor < 1 can locally/globally arrange differently under in-plane applied fields. For instance:

- $Q$-factor = 0.4: the Néel caps tended to align in the direction of the in-plane field, while most Bloch-lines aligned transverse to the in-plane field.
- $Q$-factor = 0.6: the Néel caps tended to align transverse to the in-plane field, while most Bloch-lines aligned in the direction of the in-plane field.
- $Q$-factor = 0.8: The applied in-plane field did not preferably orient the Néel caps or Bloch-line component of hybrid domain walls.

If we suppose the Fe/Gd patterned wire exhibits the same material $Q$-factor as the Fe/Gd continuous specimen ($Q$-factor ~ 0.4), we expect the domain morphology will locally/globally arrange as schematically shown in Supp. Figs. 1(a, b) under in-plane fields that are either applied parallel [Supp. Fig. 1(a)] or transverse [Supp. Fig. 1(b)] to the Fe/Gd patterned wire. Conversely, if the Fe/Gd patterned wires exhibits a material $Q$-factor = 0.6, we can expect the domain morphology will arrange as shown in Supp. Figs. 1(c, d). Since higher/lower electron scattering occurs when the magnetic spins are oriented parallel/transverse to the electrical current, we can further deduce the longitudinal resistivity response to be of the form: $\rho_{xx}^{\parallel} > \rho_{xx}^{\perp}$ [Supp. Figs. 1(a, b)] and $\rho_{xx}^{\parallel} < \rho_{xx}^{\perp}$ [Supp. Figs. 1(c, d)]. Based on the experimental temperature- and field-dependent ($\rho_{xx}^{\parallel}, \rho_{xx}^{\perp}$) longitudinal resistivity measurements, shown in Fig. 2 (main text), we know the Fe/Gd patterned wire exhibits $\rho_{xx}^{\parallel} > \rho_{xx}^{\perp}$ (at magnetic saturation) which allows us to infer the Néel caps tend to align in the direction of the in-plane field, while Bloch-lines and perpendicular stripes tend to align transverse to the in-plane field.

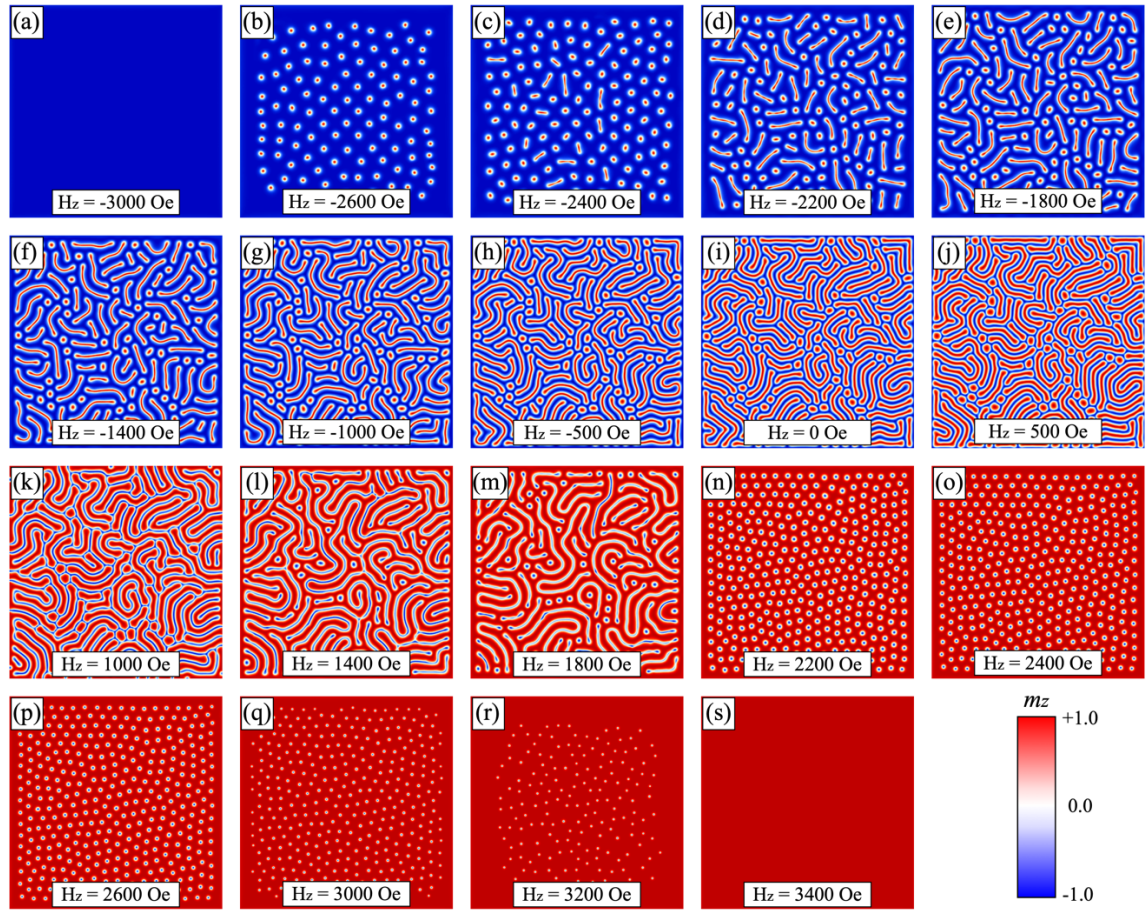

Supplementary Figure 2. Field-dependent evolution of equilibrium states as perpendicular field is swept from negative to positive saturation. Each image shows the top-side view of the magnetization along the z-axis ($m_z$) at the top surface of the slab ($z = 40\ nm$). Each image represents a 10 µm x 10 µm total area. The colorbar shows for respective polarity of the $m_z$ magnetization detailed in the equilibrium states.

## 2. Micromagnetic modeling of domain morphology.

Supplementary Figure 2 shows the field-dependent evolution of the equilibrium states that form on a 10 µm x 10 µm x 80 nm slab (with material properties: $M_S = 400$ emu/cm³, $K_U = 4 \times 10^5$ erg/cm³ and $A = 5 \times 10^{-7}$ erg/cm) as the magnetic field is swept from $H_z = -5000$ Oe to $H_z = +5000$ Oe in $H_z = 50$ Oe field steps with 10 ns rise and 20 ns relaxation time. Each image in Supp. Fig. 1 depicts the magnetization along the z-axis ($m_z$) at the top surface of the slab ($z = +40$ nm) for selected magnetic fields, which capture the evolution of magnetic phases as the perpendicular field is swept from negative to positive saturation. At $H_z = -3000$ Oe [Supp. Fig. 1(a)], the equilibrium state only exhibits $-m_z$ magnetization, which depicts the uniform phase under negative applied fields. When the field is reduced to $H_z$ = -2600 Oe [Supp. Fig. 2(b)], the equilibrium state quickly becomes populated with cylindrical spin textures that arrange in a closed packed lattice towards the center of the slab. From the micromagnetic results presented in the main text (Fig. 5), which depict the $m_x$ magnetization along the center of the slab, we know these cylindrical spin textures primarily consists of skyrmions (S= +1) with a near equal population of the two possible helicities ($\gamma = \pm\pi/2$), as well as several bubbles ($S = 0$) and antiskyrmions ($S = -1$). Further reducing the field to $H_z$ = -2400 Oe [Supp. Fig. 2(c)], the equilibrium state continues to primarily exhibit a closed-packed skyrmion lattice and we can identify several skyrmions, which begin to undergo a deformation into dumbbell-like spin textures that eventually elongate into stripes as the field is further reduced. From $H_z$ = -2400 Oe to $H_z$ = -1800 Oe [Supp. Figs. 2(c-e)], the equilibrium state shows that the stripes grow and occupy more space on the slab, whereas the number skyrmions that undergo a skyrmion-to-stripe transformation does not significantly change. Then, as the field continues to be reduced, from $H_z$ = -1800 Oe to $H_z$ = -1000 Oe [Supp. Figs. 2(e-g)], the entire slab becomes filled with a mixture of stripes and skyrmions and there is sparse space for additional magnetic spin textures. Because of the edge effects, the equilibrium states depicting a stripe phase also contain countable skyrmions because these are unable to undergo a transformation into stripes. We identify the stripe phase commences when the number of magnetic spin textures remains unchanged and only their relative size changes (i.e., under $-H_z$ fields magnetic spin textures with $-m_z$ magnetization (blue-hue-colored features) tend to be larger than those with $+m_z$ magnetization (red-hue-colored features), and vice-versa observations are noted under $+H_z$ fields), which spans from $H_z$ = -1200 Oe to $H_z$ = +800 Oe [Supp. Figs 2(g-j)]. Note that the stripe phase is not symmetric around the zero-field. When increasing the field to $H_z$ = +1000 Oe [Supp. Figs 2(k)], we identify that several skyrmions merge with stripes in their local vicinity. Also, when comparing the equilibrium states that form under $H_z$ = ±1000 Oe [Supp. Figs. 2(g, k)], we clearly discern that a difference in the volume fraction (at the top surface of the slab) exists between the opposite $m_z$ magnetization spin textures (blue- and red-hue-colored features) and their domain wall (white-hue-colored features). When further increasing the field, from $H_z$ = +1000 Oe to $H_z$ = +1800 Oe [Supp. Figs. 2(k-m)], the stripes begin to collapse into cylindrical spin textures while remaining stripes exhibit a domain wall width that increases and extends over the $-m_z$ magnetization of the stripe; the latter can clearly be noted when comparing equilibrium states that form under $H_z = \pm1800$ Oe [Supp. Figs. 2(e, m)]. These changes in the surface-component of the Néel caps match the applied fields, where the Bloch-line chirality of stripes also undergoes local changes [Fig. 5(p-r)]. At $H_z = +2200$ Oe [Supp. Fig. 2(n)], the equilibrium state consists of a high-density closed-packed lattice of skyrmions, which we know from inspecting the Bloch-line chirality along the center of the slab [Fig. 5, main text]. As the field is increased to positive magnetic saturation [Supp. Fig. 2(n-q)], we find that the closed-packed skyrmion lattice phase persists over a larger field range under positive fields (i.e., $H_z = +2200$ Oe to $H_z = +3200$ Oe) as compared to the one formed under negative applied fields (i.e., from $H_z = -2600$ Oe to $H_z = -2450$ Oe). As the field approaches the magnetization saturation, a disordered skyrmion phase is only observed from $H_z = +3200$ Oe to $H_z = +3350$ Oe [Supp. Fig. 2(r)]. At/above $H_z = +3200$ Oe [Supp. Fig. 2(s)], the equilibrium state shows a uniform phase with only $+m_z$ magnetization.

3. *Field-dependent evolution ground state with global achiral phase.*

Supplementary Figure 3 schematically shows a potential achiral ground state and its anticipated Bloch-line evolution under applied fields for a dipole magnet hosting a skyrmion lattice phase under negative and positive applied fields. Since we know the skyrmion lattice phase primarily consists an equal population of 50% left-handed (*L*) and 50% right-handed (*R*) skyrmions [Supp. Fig. 3(a)], then reducing the field towards zero can result in formation of $+m_z$ magnetization stripes with 50% *L*- and 50% *R*-Bloch helicity and $-m_z$ magnetization stripes with 50% *L*- and 50% *A*-Bloch helicity [Supp. Fig. 3(b, c)]. As noted in the main text, the Bloch-line chirality undergoes local changes to support the formation of a skyrmion lattice under positive perpendicular fields, which also consists of an equal population of skyrmions with 50% *L*- and 50% *R*-Bloch helicity. Unlike the case shown in Figure 6(f-j), main text, where the minimal Bloch-line rotations are required to achieve the skyrmion lattice phase, if the domain morphology now undergoes 4 Bloch-line rotations [Supp. Fig. 3(d)] then it is possible to achieve the desired skyrmion lattice [Supp. Fig. 3(e)]. Since the domain morphology of a magnet results from the minimization of the total magnetic energy, we anticipate that the potential field-driven Bloch-line rearrangement depicted in Supp. Fig. 3 is energetically unfavorable because unlike the case illustrated in Fig. 6(f-j) [main text] three additional Bloch-line are required.

Supplementary Figure 4 schematically shows an alternate ground state and its anticipated Bloch-line evolution under applied fields for a dipole magnet, which can also faciliate the formation of a skyrmion lattice under negative and positive applied fields. Unlike the ground state illustrated in Fig. 6(f-j) [main text] and Supp. Fig. 4, two Bloch-line rotations are required to achieve a skyrmion lattice phase under positive applied fields which can be achieved via two different local Bloch-line rearrangements [Supp. Figs. 4(a-e) and 4(f-j)]; the main distinction is that in one case the total chirality (i.e., the one accounting the Block-line chirality of $+m_z$ and $-m_z$ magnetic spin textures under positive and negative fields) is not conserved [Supp. Fig. 4(a-e)] while in the other scenario it is [Supp. Fig. 4(f-j)].

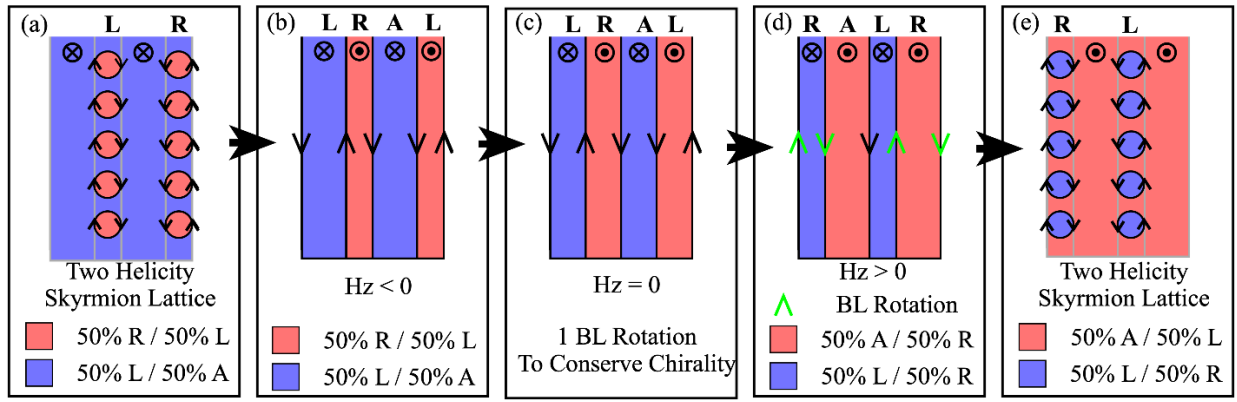

Supplementary Figure 3. Schematic illustration of the Bloch-line evolution for a global achiral skyrmion lattice phases as the perpendicular magnetic field is applied from negative to positive (left to right images) saturation. The right-handed (R), left-handed (L), and achiral (A) Bloch-line helicity is depicted with a black-hue arrowheads along the perimeter of red/blue-colored features which represent positive/negative perpendicular magnetic domains. A green-hue arrow head depicts a Bloch-line whose helicity changed compared to prior applied fields. For each spin texture arrangement, the Bloch-line helicity (L, R, A) distribution is depicted for both $\pm m_z$ magnetization textures.

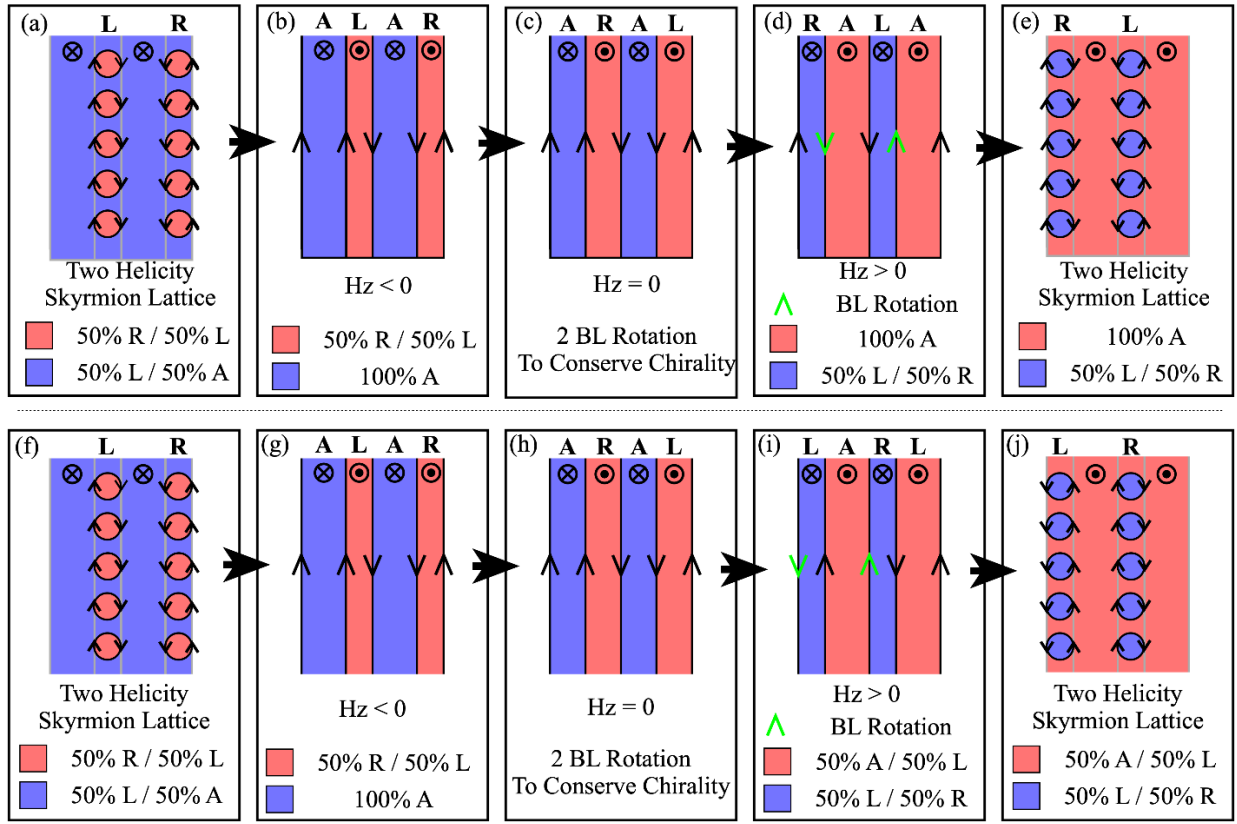

Supplementary Figure 4. Schematic illustration of the Bloch-line evolution global achiral skyrmion lattice phases as the perpendicular magnetic field is applied from negative to positive (left to right images) saturation. The initial ground state (a-b, f-g) can result in two (a-e, and f-j) distinct Bloch-line evolution paths in which different Bloch-line rotate under applied fields to achieve the equal-population global achiral skyrmion lattice phase. The right-handed (R), left-handed (L), and achiral (A) Bloch-line helicity is depicted with a black-hue arrowheads along the perimeter of red/blue-colored features which represent positive/negative perpendicular magnetic domains. A green-hue arrowhead depicts a Bloch-line whose helicity changed compared to prior applied fields. For each spin texture arrangement, the Bloch-line helicity (L, R, A) distribution is depicted for both $\pm m_z$ magnetization textures.

References.